\documentclass[12pt]{article}

\global\arraycolsep=1pt
\usepackage{amssymb}
\usepackage {pstricks}
\newcommand{\beq}{\begin{equation}}
\newcommand{\eeq}{\end{equation}}
\newcommand{\beqa}{\begin{eqnarray}}
\newcommand{\eeqa}{\end{eqnarray}}
\newcommand{\CR}{\nonumber \\}
\newcommand{\vect}{\overrightarrow}
\newcommand{\lam}{\lambda}
\newcommand{\m}{\mu}

\renewcommand{\k}{\kappa}

\renewcommand{\theequation}{\thesection.\arabic{equation}}
\renewcommand{\thefootnote}{\fnsymbol{footnote}}
\newcommand{\unitbox}
{\setlength{\unitlength}{0.5pt}
\begin{picture}(10,10)
\put(0,10){\line(1,0){10}}
\put(0,0){\line(1,0){10}}
\put(0,0){\line(0,1){10}}
\put(10,0){\line(0,1){10}}
\end{picture}}

\begin{document}

\begin{titlepage}
\begin{flushright}
{February, 2005} \\
{\tt hep-th/0502061} 
\end{flushright}
\vspace{0.5cm}
\begin{center}
{\Large \bf
Instanton counting, Macdonald function \\
and the moduli space of 
$D$-branes}
\vskip1.0cm
{\large Hidetoshi Awata and Hiroaki Kanno}
\vskip 1.0em
{\it 
Graduate School of Mathematics \\
Nagoya University, Nagoya, 464-8602, Japan}
\end{center}
\vskip1.5cm
%%%%%%

%%%%%%%%%

\begin{abstract}

We argue the connection of Nekrasov's partition function 
in the $\Omega$ background and the moduli space of $D$-branes,
suggested by the idea of geometric engineering and Gopakumar-Vafa
invariants. In the instanton expansion of ${\cal N} =2$ $SU(2)$ Yang-Mills theory
the Nakrasov's partition function with equivariant parameters 
$\epsilon_1, \epsilon_2$ of toric action on ${\mathbb C}^2$ factorizes 
correctly as the character of $SU(2)_L \times SU(2)_R$ spin representation.
We show that up to two instantons the spin contents are consistent
with the Lefschetz action on the moduli space of $D2$-branes on (local) ${\bf F}_0$.
We also present an attempt at constructing 
a refined topological vertex in terms of the Macdonald function. 
The refined topological vertex with two parameters of $T^2$ action allows us
to obtain  the generating functions of equivariant $\chi_y$
and elliptic genera of the Hilbert scheme of $n$ points on ${\mathbb C}^2$
by the method of  topological vertex. 
%\vspace{-3.0mm}
%\begin{flushleft}
%Keywords: 
%\end{flushleft}
\end{abstract}
\end{titlepage}

%%%%%%%%%%%%%%%%%%%%%%%%%%%%%%%%%%%%%%%%%%%%%%%%%%%%%%

\renewcommand{\thefootnote}{\arabic{footnote}}
\setcounter{footnote}{0}

%%%%%%%%%%%%%%%%%%%%%%%%%%%%%%%%%%%%%%%%%%%%%%%%%%%%%%%

\section{Introduction}
\setcounter{equation}{0}

Gravity/gauge theory correspondence is one of important subjects in the recent
developments of non-perturbative string theory and $D$-branes are the 
crucial dynamical objects for understanding the correspondence. 
In topological theory the correspondence can be established more precisely 
as equalities among
partition functions of topological string/gauge theory. Since the topological
partition function gives the generating function of topological invariants,
the equalities imply rather surprising mathematical conjectures on the relation 
of invariants that seem to have quite different origin at the first sight. 
The first example was given by the geometric transition based on the duality 
of the resolved conifold and the deformed conifold \cite{GV1}\cite{GV2}. 
In $A$-model picture the geometric transition
implies the equivalence of the Gromov-Witten invariants of local ${\bf P}^1$
(the resolved conifold side) and the Chern-Simons invariants of $S^3$ (the
deformed conifold side), while in $B$-model picture it leads to the Dijkgraaf-Vafa
proposal of matrix model computation of effective superpotential of ${\cal N}=1$
supersymmetric Yang-Mills (SYM) theory \cite{DV}. 
Thanks to a technical tool of topological vertex arising from the idea of 
geometric transition \cite{Iqb}\cite{DFG}\cite{AMV}\cite{AKMV}\cite{LLLZ}, 
more examples of topological string/gauge theory correspondence have been 
found in the last years. Among them are the equivalence of the Nekrasov's partition function
for instanton counting of four dimensional gauge theory and topological string amplitudes 
of local toric Calabi-Yau manifold \cite{IK-P1}\cite{IK-P2}\cite{EK1}\cite{EK2}\cite{Zhou1}, 
and Gromov-Witten/Donaldson-Thomas correspondence \cite{ORV}\cite{INOV}\cite{MNOP1}\cite{MNOP2}.
Regarding these topological theories as one of the \lq\lq corners\rq\rq\ of a hypothetical 
topological theory like the picture of $M$ theory for the duality web of perturbative string theories,
there have been a few proposal of unifying topological theory \cite{DGNV}\cite{NekZ}
that accommodates all the correspondences and  encodes the idea of topological $S$-duality \cite{NOV}\cite{Kap}.¡¡

There is a rather long history of attempts to recover the instanton expansion of the Seiberg-Witten theory 
from the calculus of multi-instantons (see \cite{DHKM} and references thererin).
The main issue was the measure on the instanton moduli space based on the ADHM construction.
The topological nature of the instanton amplitudes was understood from the existence of 
a BRST-type operator \cite{BFTT} and consequently the localization theorem was successfully applied \cite{Hol}. 
In the end Nekrasov correctly computed the partition function 
that produced the instanton expansion of the Seiberg-Witten prepotential
by putting ${\cal N} =2$ SYM in the $\Omega$ background \cite{NekSW}.
The $\Omega$ background has a natural interpretation from the six dimensional
viewpoint and it gives a coupling with gravity in a special form.
The background has two parameters $\epsilon_1$ and $\epsilon_2$,
which define a twisting of two complex coordinates $(z_1, z_2)$
of the flat (Euclidean) space ${\mathbb R}^4 \simeq {\mathbb C}^2$;
\beq
(z_1, z_2) \mapsto (e^{i\epsilon_1} z_1, e^{i\epsilon_2} z_2)~.
\eeq
Mathematically they are equivariant parameters corresponding to the weights
of $T= {\mathbb C}^\times \times  {\mathbb C}^\times$ action on ${\mathbb C}^2$. 
For any manifold $X$ with $T$ action the equivariant cohomology ring 
$H^*_T(X,{\mathbb Q})$ is a module over $H^*_T({\rm pt}) \simeq {\mathbb Q}[t,q],~
t:=e^{\epsilon_1}, q:=e^{-\epsilon_2}$.
When we consider the special case $\hbar := \epsilon_1 = - \epsilon_2$, 
the Nekrasov's partition function has the following expansion
\beq
Z_{Nek} = \exp \left( \sum_{g \geq 0} \hbar^{2g-2} F_g \right)~,
\eeq
and the lowest term $F_0$ gives the instanton part of the Seiberg-Witten
prepotential \cite{NO}\cite{NY1}\cite{NY2}. 
The fact that $Z_{Nek}$ can be expanded like the genus expansion
in string theory is not an accident and it turns out $Z_{Nek}$ is exactly reproduced
as a (closed) topological string amplitudes on an appropriate background.
The amplitudes of our concern are those on local toric Calabi-Yau manifolds and
the computation is facilitated by making use of topological vertex introduced in \cite{AKMV}.
The topological vertex is related to the Hopf link invariants of
the Chern-Simons theory in large $N$ limit and
they are given by (the special values of)
the Schur function \cite{ML}. In the theory of symmetric functions,
a generalization of the Schur function with two parameters is known and
it is called Macdonald function \cite{Mac}. Then it is natural to expect
that the Nekrasov's partition function in general $\Omega$ background is
expressed in terms of the Macdonald function. 
The aim of this article is to explore the relation of the Macdonald function
to the Nakrasov's partition function. 

The counting of BPS states in five dimensional theories is 
one of natural ways to understand the equivalence of the Nekrasov's partition 
function for instanton counting in four dimensions and topological string amplitudes 
which are given by Gopakumar-Vafa invariants \cite{GV3}\cite{GV4}. 
In the space-time interpretation of topological string amplitudes
we consider a constant self-dual graviphoton background and this is 
in fact what Nekrasov did in his $\Omega$ background 
with the constraint $\epsilon_1  + \epsilon_2 =0$. 
In section two we consider Nekrasov's formula for the $SU(2)$ gauge theory and
check that, even if $\epsilon_1  + \epsilon_2 \neq 0$ and
the self-duality is lost,  it factorizes correctly as the character 
of $SU(2)_L \times SU(2)_R$ spin representation of five dimensional massive particles.
Part of similar results was already reported in \cite{HIV} and we will extend it up to
two instantons. The idea of geometric engineering tells us that $SU(2)$ gauge theory can be
geometrically engineered by type IIA compactification on local Calabi Yau space
$K_{{\bf F}_0}$; the canonical bundle of the Hirzebruch surface ${\bf F}_0 \simeq
{\bf P}^1 \times {\bf P}^1$. In the attempts at a mathematical definition 
of the Gopakumar-Vafa invariants, it has been argued that the spin contents
of five dimensional theory are identified with the Lefschetz decomposition of the
cohomology ring of the moduli space of $D2$-branes \cite{KKV} \cite{HST}. 
We find that the spin contents obtained by Nekrasov's  formula are consistent
with this interpretation of Gopakumar-Vafa invariants on local ${\bf F}_0$.

The fact that Nekrasov's partition function in general $\Omega$ background 
allows a factorization in terms of the spectrum of five dimensional massive particles
strongly suggests the existence of topological vertex with two parameters.
In section three we make a modest step towards constructing such a refined topological
vertex. We note that a similar refinement in the context of open string theory 
and its relation to knot invariants have been discussed quite recently \cite{GSV}.
We propose the refined topological vertex in terms of the specialization of
the Macdonald function that is a two parameter generalization of the Schur function.
As a preliminary example we consider $U(1)$ gauge theory with a 
massive adjoint hypermultiplet, which is mathematically
related to the Hilbert scheme ${\rm Hilb}^n~{\mathbb C}^2$ 
of $n$ points on ${\mathbb C}^2$.
We show that the generating functions of equivariant $\chi_y$ genera and 
elliptic genera of ${\rm Hilb}^n~{\mathbb C}^2$ are obtained by the method of
topological vertex. 
Several technical points on partitions and Macdonald functions are summarized
in appendices. 

We have checked that Nekrasov's formula for $SU(N)$ gauge theory in general
$\Omega$ background is reproduced by the refined topological vertex 
proposed in this paper. However, we have the issue of cyclic symmetry of the vertex.
Therefore, the amplitude obtained by the method of topological vertex may not be
unique for a given toric diagram. The results will be reported elsewhere \cite{AK}.

%%%%%%%%%%%%%%%%%%%%%%%%%%%%%%%%%%%%%%%%%%%%%%%%%%%%%%%

\section{Nekrasov's partition function and the degeneracy of BPS states}
\setcounter{equation}{0}

%%%%%%%%%%%%%%%%%%%%%%%%%%%%%%%%%%%%%%%%%%%%%%%%%%%%%%%

The $\Omega$-background of Nekrasov has two parameters 
$(\epsilon_1, \epsilon_2)$ and introduces physically 
a constant graviphoton field of ${\cal N}=2$ supergravity in four dimensions
\footnote{In fact this point becomes more clear, if the theory is lifted 
to five dimensions \cite{Tac}.}.
The condition $\epsilon_1 + \epsilon_2 =0$ implies that
the graviphoton background is self-dual and thus satisfies the BPS condition. 
A constant self-dual graviphoton background plays quite a significant role 
in the space-time interpretation of topological string amplitudes
and it is natural that we can relate the Nekrasov's partition function 
to topological string amplitudes, when $\epsilon_1 + \epsilon_2 =0$.
It gives an $A$-model amplitude that only depends on the K\"ahler parameters
and is invariant under the deformation of complex structure of the target.

In a general $\Omega$-background, where we have no longer $\epsilon_1 + \epsilon_2 =0$,
the background breaks the BPS condition. But due to their topological nature,
we expect that the instanton amplitudes in four dimensional gauge theory
are unaffected by the deformations. 
From the viewpoint of topological $A$-model, one may worry about the fact that
the independence of the amplitudes under the deformation is not protected by supersymmetry.  
However, as has been pointed out in \cite{HIV}, when we consider non-compact Calabi-Yau 
manifolds\footnote{This is the case in the geometric engineering
of four dimensional gauge thoery.}, there are no (well-defined)
deformations of complex structure and we do not have to mind the jump
of the spectrum of the BPS states. In other words it is not required to take
a supersymmetric trace with respect to the spin contents of $SU(2)_R$ action.

%%%%%%%%%%%%%%%%%%%%%%%%%%%%%%%%%%%%%%%%%%%%%%%%%%%%%
\subsection{Moduli space of $D2$-branes}

Let us review the relation of (generalized) Gopakumar-Vafa invariants\footnote{For recent
developments in Gopakumar-Vafa invariants, see \cite{Peng} \cite{Kon} \cite{SS}.}  and
the Lefschetz action on the cohomology of the moduli space $\widetilde{\cal M}_\beta$
of $D2$-branes wrapping on the holomorphic two-cycle 
$\beta \in H_2(X, {\mathbb Z})$.  The cohomology classes in $H^*(\widetilde{\cal M}_\beta)$ are
in one to one correspondence with the BPS states in a five dimensional theory
obtained by a compactification of $M$-theory on a Calabi-Yau 3-fold $X$.
These BPS states are labeled by the representation of the spacial
rotation group $SO(4) \simeq SU(2)_L \times SU(2)_R$ in five dimensions 
and the degeneracy of the BPS states can be written as 
\beq
\left[ \left( \frac{1}{2}, 0 \right) \oplus 2(0,0) \right] \otimes \bigoplus_{(j_L, j_R)}
N_\beta^{(j_L, j_R)} \left[ (j_L, j_R) \right]~.
\eeq
Since the BPS state preserves half the supersymmetry, we always have the 
structure of the half hypermultiplet $\left[ \left( \frac{1}{2}, 0 \right) \oplus 2(0,0) \right]$.
The integers $N_\beta^{(j_L, j_R)}$ denote the number of BPS states
with the central charge of the homology class $\beta$ and $SU(2)_L \times SU(2)_R$
spin $(j_L, j_R)$. The moduli space of $D2$-branes consists of the deformation 
of the holomorphic cycle in $X$ together with the moduli of flat (=stable) 
$U(1)$ bundle over it. Thus we have a fibration $\pi : \widetilde{\cal M}_\beta \to {\cal M}_\beta$,
where the base ${\cal M}_\beta$ is the moduli space of the two-cycle $\beta$ without
a choice of flat bundle. If the two-cycle is generically the Riemann surface of genus
$g$, then the generic fiber is $T^{2g}$, the Jacobian variety of the Riemann surface. 
Both $\widetilde{\cal M}_\beta$ and ${\cal M}_\beta$ are K\"ahler manifolds and 
we have the Lefschetz action on the cohomology group defined by the multiplication of a K\"ahler form.
It has been argued that the $SU(2)_L$ spin is identified with the Lefschetz
decomposition along the fiber direction (a relative Lefschetz action) and the $SU(2)_R$ 
corresponds to the action on the base space \cite{KKV} \cite{HST}. 
Thus we have the following decomposition of
the cohomology of the moduli space of $D2$ branes;
\beq
H^*(\widetilde{\cal M}_\beta) = \sum N_\beta^{(j_1, j_2)} \left[ (j_1^{fiber}, j_2^{base})  \right]~.
\eeq
In particular this identification implies that the $SU(2)_R$ spin contents with the highest $SU(2)_L$
spin are given by the Lefschetz decomposition of the cohomology of the base 
space ${\cal M}_\beta$.

We thus expect the Nekrasov's partition function in a general $\Omega$
background should be related to 
the $SU(2)_L \times SU(2)_R$ spin contents of BPS states obtained
by wrapping $M2$ branes on a supersymmetric(=holomorphic) two-cycle.
It is a very hard mathematical problem to work out such a Lefschetz
decomposition of $H^*(\widetilde{\cal M}_\beta)$
in general. Fortunately we have some information on this issue
for local ${\bf F}_0 = {\bf P}^1 \times {\bf P}^1$, to which the $SU(2)$ pure
Yang-Mills theory is expected to correspond. In the following we will obtain
a few predictions on the Lefschetz decomposition
from the Nekrasov's partition function for $SU(2)$ Yang-Mills theory
and show that they are consistent with the known results on the moduli space
of $D2$ branes on local ${\bf F}_0$.

In the space-time interpretation of topological string amplitudes the free energy 
$F = \log Z = \sum_{g=0}^\infty g_s^{2g-2} F_g$ is obtained by summing over
the contribution of the BPS particle spectrum with multiplicities $N_\beta^{(j_L,j_R)}$ 
to the low energy effective action. We thus obtain in general
\beqa
F(q,t;Q_\beta) &=& \sum_{\beta \in H_2(X, {\mathbb Z})} \sum_{n=1}^\infty
\sum_{j_L, j_R} \frac{N_\beta^{(j_L,j_R)}}{n(q^{n/2} - q^{-n/2})(t^{n/2} - t^{-n/2})} \CR
& & \times \left( (qt)^{-n\cdot{j_L}} + \cdots  + (qt)^{n\cdot{j_L}} \right)
\left( (q/t)^{-n\cdot{j_R}} + \cdots + (q/t)^{n\cdot{j_R}} \right) Q_\beta^n~, 
\eeqa
where we have introduced a constant graviphoton background with
$F_{12}= \epsilon_1, F_{34}= \epsilon_2$ and put $t:=e^{\epsilon_1}, 
q:=e^{-\epsilon_2}$. The first summation is over the second homology classes $\beta$
of the target space $X$ with the integer coefficients. In the BPS state counting
the two cycle $\beta$ should be supersymmetric, that means it is holomorphic.
Hence only non-negative part contributes to the sum over $H_2(X, {\mathbb Z})$. 
$Q_\beta= e^{-t_\beta}$ and $t_\beta = \int_\beta \omega$, 
with $\omega$ being a K\"ahler form of the target space,
namely $t_\beta$ is the K\"ahler parameter associated with the two cycle $\beta$. 
The summation over $n$ accounts for the multi-covering. 
Since the multiplicity $N_\beta^{(j_L,j_R)}$ is independent of $n$, we can compute it by extracting 
the $n=1$ term from the sum of the multi-covering.
It is convenient to make a change of variables; $q=uv, t=u/v$. Then the free energy is
\beqa
F(q,t;Q_\beta) &=& \sum_{\beta \in H_2(X, {\mathbb Z})} 
\sum_{j_L, j_R} \sum_{n=1}^\infty
\frac{N_\beta^{(j_L,j_R)} u^n v^n}{n(u^n v^n -1)(u^n- v^n)} \CR
& &~~\left( \frac{u^{2n \cdot  j_L +1} - u^{-2n \cdot  j_L -1}}{u - u^{-1}} \right) \cdot
\left( \frac{v^{2n \cdot  j_R +1} - v^{-2n \cdot  j_R -1}}{v - v^{-1}} \right) Q_\beta^n~. \label{GVrefined}
\eeqa
Note that 
\beq
\chi_n (q) := q^n + q^{n-2} + \cdots + q^{-n+2} + q^{-n} 
= \frac{q^{n+1} - q^{-n-1}}{q - q^{-1}}~,
\eeq
is the character of the irreducible representation of $SU(2)$ with spin $n/2$.

%%%%%%%%%%%%%%%%%%%%%%%%%%%%%%%%%%%%%%%%%%%%%%%%%%%%%%
\subsection{$SU(2)_L \times SU(2)_R$ decomposition}

In \cite{FP} \cite{BFMT} \cite{NY1} the Nekrasov's partition function for $SU(N_c)$ Yang-Mills theory
in a general $\Omega$ background is written in the following form;
\beq
Z(\epsilon_1, \epsilon_2, \vect{a}, \Lambda)
= \sum_{\vect{Y}} \frac{\Lambda^{|\vect{Y}|}}
{\prod_{\alpha, \beta} n_{\alpha, \beta}^{\vect{Y}}
(\epsilon_1, \epsilon_2, \vect{a})}~, \label{NY}
\eeq
where
\beq
n_{\alpha, \beta}^{\vect{Y}} = \prod_{s \in Y_\alpha}\left(-\ell_{Y_\beta}(s) \epsilon_1
+ (a_{Y_\alpha}(s) + 1) \epsilon_2 + a_\beta - a_\alpha\right)
\prod_{t \in Y_\beta}\left((\ell_{Y_\alpha}(t) +1) \epsilon_1
- a_{Y_\beta}(t) \epsilon_2 + a_\beta - a_\alpha\right)~.
\eeq
We follow the notations in \cite{NY1}. 
The right hand side of (\ref{NY}) is a summation over $N_c$-tuples of 
Young diagrams $\vect{Y} = (Y_1, Y_2, \cdots, Y_{N_c})$ 
or colored partitions\footnote{In this paper
we often identify Young diagrams, partitions and representations of general linear group.}.
The total number of boxes of $\vect{Y}$ is denoted by $|\vect{Y}|$ and identified
with the instanton number. $\ell_Y(s)$ and $a_Y(s)$ are the leg-length and the arm-length
of the Young diagram (see Appendix A.1). In deriving (\ref{NY}) an equivariant integration
over the (framed) instanton moduli space ${\cal M}_{ADHM}$ has to be computed and
the localization principle is employed. The $N_c$-tuples of Young diagrams $\vect{Y}$ are
in one to one correspondence with the fixed points of the toric action on ${\cal M}_{ADHM}$.
The denominator $\prod_{\alpha, \beta} n_{\alpha, \beta}^{\vect{Y}} $ is 
the product of the eigenvalues (weights) of the toric action at the fixed point $\vect{Y}$. 
Note that the number of the factors in 
$\prod_{\alpha, \beta} n_{\alpha, \beta}^{\vect{Y}} $ is $2 k N_c$ and
this is nothing but the (complex) dimensions of ${\cal M}_{ADHM}$.

Introducing the notations $t := e^{\epsilon_1}, q := e^{-\epsilon_2}$ and $Q_{\beta\alpha}
:= e^{a_\beta - a_\alpha}$, we consider the following five dimensional lift;
\beq
N_{\alpha, \beta}^{\vect{Y}} = \prod_{s \in Y_\alpha}
\left( 1-t^{-\ell_{Y_\beta}(s)} q^{- (a_{Y_\alpha}(s) + 1)} Q_{\beta\alpha} \right)
\prod_{t \in Y_\beta}\left( 1- t^{\ell_{Y_\alpha}(t) +1} 
q^{a_{Y_\beta}(t)} Q_{\beta\alpha} \right)~. \label{5Dlift}
\eeq
More explicitly for $SU(2)$ case $(\vect{Y} = (Y_1, Y_2)$ and $Q := Q_{12} = Q_{21}^{-1}$);
we have
\beqa
\prod_{\alpha, \beta} N_{\alpha, \beta}^{(Y_1, Y_2)} &=& 
N_{11}^{(Y_1, Y_2)} N_{22}^{(Y_1, Y_2)} N_{12}^{(Y_1, Y_2)} N_{21}^{(Y_1, Y_2)}  \CR
&=&\prod_{s \in Y_1}\left( 1-t^{-\ell_{Y_1}(s)} q^{- (a_{Y_1}(s) + 1)} \right)
\left( 1- t^{\ell_{Y_1}(s) +1} q^{a_{Y_1}(s)} \right) \CR
& &~\prod_{t \in Y_2}\left( 1-t^{-\ell_{Y_2}(t)} q^{- (a_{Y_2}(t) + 1)} \right)
\left( 1- t^{\ell_{Y_2}(t) +1} q^{a_{Y_2}(t)} \right) \CR
& &~\prod_{s \in Y_1}\left( 1-t^{-\ell_{Y_2}(s)} q^{- (a_{Y_1}(s) + 1)} Q^{-1}\right)
\prod_{t \in Y_2}\left( 1- t^{\ell_{Y_1}(t) +1} q^{a_{Y_2}(t)} Q^{-1}\right) \CR
& &~\prod_{s \in Y_2}\left( 1-t^{-\ell_{Y_1}(s)} q^{- (a_{Y_2}(s) + 1)} Q \right)
\prod_{t \in Y_1}\left( 1- t^{\ell_{Y_2}(t) +1} q^{a_{Y_1}(t)} Q \right).
\eeqa
We note the following symmetry;
\beqa
\prod_{\alpha, \beta} N_{\alpha, \beta}^{{(Y_1, Y_2)}}(q,t,Q) 
&=& \prod_{\alpha, \beta} N_{\alpha, \beta}^{{(Y_1^t, Y_2^t)}}(t^{-1}, q^{-1},Q)~,  \\
\prod_{\alpha, \beta} N_{\alpha, \beta}^{{(Y_1, Y_2)}}(q,t,Q) 
&=& \prod_{\alpha, \beta} N_{\alpha, \beta}^{{(Y_2, Y_1)}}(q, t,Q^{-1})~, 
\eeqa
where $Y^t$ means the transpose of the Young diagram. 
Since these appear in pairs in the sum with a fixed instanton number,
the final expression should be symmetric under $(q,t) \leftrightarrow (t^{-1}, q^{-1})$ 
and $Q \leftrightarrow Q^{-1}$.

Let us look at examples at lower instanton numbers.
At one instanton we have either $\vect{Y}=(\unitbox, \bullet)$ or 
$\vect{Y}=(\bullet, \unitbox)$ and the partition function is
\beqa
Z^{\hbox{one-inst}} (q,t, Q) &=& \frac{1}{(1- q^{-1})(1-t)(1-tq^{-1} Q^{-1})(1-Q)}  \CR
& &~~~+\frac{1}{(1- q^{-1})(1-t)(1- Q^{-1})(1- t q^{-1}Q)}~. \label{1inst}
\eeqa
There are five possibilities of $(Y_1, Y_2)$ for two instanton 
and we obtain;
\beqa
Z^{\hbox{two-inst}} (q,t, Q) &=&
\frac{1}{(1-t^{-1} q^{-1})(1-q^{-1})(1-t^2)(1-t)} \CR
& &\times \left[ \frac{1}{(1-t q^{-1} Q^{-1})(1-t^2 q^{-1} Q^{-1})(1-Q)(1-t^{-1}Q)} \right.\CR
& &~~~+ \left.\frac{1}{(1-t q^{-1} Q)(1-t^2 q^{-1} Q)(1-Q^{-1})(1-t^{-1} Q^{-1})} \right]  \CR
& & + \frac{1}{(1- q^{-2})(1-q^{-1})(1-t q)(1-t)} \label{2inst} \\
& &\times \left[ \frac{1}{(1-t q^{-2} Q^{-1})(1-t q^{-1} Q^{-1})(1- q Q)(1- Q)} \right.\CR
& &~~~+ \left.\frac{1}{(1-t q^{-2} Q)(1-t q^{-1} Q)(1- q Q^{-1})(1- Q^{-1})} \right] \CR
& & + \frac{1}{(1-q^{-1})^2 (1-t)^2} \frac{1}{(1 - q^{-1} Q^{-1})(1-t Q^{-1})
(1 - q^{-1} Q )(1-t Q)}~. \nonumber
\eeqa
It is convenient to use the variable $v^2 := q/t$ and rewrite the one-instanton part
of the partition function as follows;
\beq
(q^{1/2} - q^{-1/2})(t^{1/2} - t^{-1/2}) Z^{\hbox{one-inst}}(q,t,Q)
= \frac{ v Q}{1-Q} \left(\frac{v^2}{1- v^2Q} + \frac{1}{1- v^{-2} Q} \right)~.
\eeq
We define the prepotential by
\beq
F := \log Z = \log ( 1 + \Lambda \cdot Z^{\hbox{one-inst}} + \Lambda^2 
\cdot Z^{\hbox{two-inst}} + \cdots)~.
\eeq
Then we have the following expansion of the one instanton contribution
to the prepotential $F^{\hbox {one-inst}} =Z^{\hbox{one-inst}}$;
\beqa
F^{\hbox{one-inst}} (q,t,Q) &=& \frac{v^2 Q}{(q^{1/2} - q^{-1/2})(t^{1/2} - t^{-1/2})}
\left( \sum_{n=0}^\infty Q^n \right) \left( \sum_{k=0}^\infty (v^{2k+1} + v^{-2k -1} ) Q^k \right) \CR
&=& \frac{v^2 Q}{(q^{1/2} - q^{-1/2})(t^{1/2} - t^{-1/2})}
\sum_{n=0}^\infty \left( \sum_{k=0}^n (v^{2k+1} + v^{-2k -1} ) \right) Q^n~.
\eeqa
Comparing this expansion with the general form (\ref{GVrefined}) of the free energy,
we find
\beq
N_{B + n F}^{(j_L, j_R)} = \delta_{j_L, 0} \delta_{j_R, n+\frac{1}{2}}~,
\eeq
and that one instanton part is saturated by the $j_L =0$ contributions.
As has been argued in \cite{HIV}, this result is consistent with the geometry of
the moduli space of $D2$ branes. It is known that
a curve of bi-degree $(a,b)$ in ${\bf P}^1 \times {\bf P}^1$ 
has total degree $d=a+b$ and generically genus $g=(a-1)(b-1)$.
Furthermore the moduli space of such curves (without the flat bundle over them)
is shown to be ${\bf P}^{(a+1)(b+1)-1}$. When $a=1$ the curve has generically
genus zero and the fiber is trivial, implying there is no non-vanishing left spin.
The moduli space of the curve is  ${\bf P}^{2n+1}$ and the Lefschetz decomposition
of $H^*({\bf P}^{2n+1})$ gives a single multiplet of spin $n + 1/2$.

The prepotential of two instanton part is given by
\beq
F^{\hbox{two-inst}}(q,t,Q) = Z^{\hbox{two-inst}}(q,t,Q) 
 - \frac{1}{2} (Z^{\hbox{one-inst}}(q,t,Q))^2~,
\eeq
and the effect of multicovering is subtracted by 
\beq
{\widetilde F}^{\hbox{two-inst}}(q,t,Q) 
= F^{\hbox{two-inst}}(q,t,Q)-\frac{1}{2} Z^{\hbox{one-inst}}(q^2, t^2, Q^2)~.
\eeq
From (\ref{1inst}) and (\ref{2inst}) we obtain
\beq
{\widetilde F}^{\hbox{two-inst}}(u,v,Q) = \frac{uv}{(uv-1)(u-v)}
\frac{N(u,v,Q)}{D(u,v,Q)} \cdot (v^2Q)^2~,
\eeq
where 
\beqa
D(u,v,Q) &=& \left (1- v^2 Q\right )\left (1- v^4 Q^2\right )\left (1- uv^3 Q\right )
\left (1 - u^{-1} v^3 Q\right ) \CR
& & \times \left (1 - v^{-2} Q\right )\left (1 - v^{-4} Q^2\right )
\left (1 - u v^{-3} Q\right )\left (1 - u^{-1} v^{-3} Q\right )~,
\eeqa 
and
\beqa
N(q,t,Q)&=& Q^3 \left[
\left (v^5 + v^3 + v + v^{-1} + v^{-3} + v^{-5} \right )
\left( {Q}^{2} + {Q}^{-2} \right)  \right.\CR
& & +\left ( v^5 + v^{-5} - (v^2 + 1 + v^{-2})(u + u^{-1})\right ) 
\left( {Q} + {Q}^{-1} \right) \CR
& & \left. +\left ( v^5 - v - v^{-1} + v^{-5} - (v^2 + v^{-2})(u+u^{-1})
\right ) \right]~.
\eeqa
The expansion of ${\widetilde F}^{\hbox{two-inst}}$ with respect to the parameter $Q$ gives
\beq
{\widetilde F}^{\hbox{two-inst}}(u,v,Q) = \frac{uv^5 Q^2}{(uv-1)(u-v)} 
\frac{1}{(u-u^{-1})(v - v^{-1})} \sum_{k=1}^\infty GV_{2,k} Q^k~,
\eeq
where
\beq
GV_{2,k} = \sum_{\ell=1}^k U_\ell \sum_{m=1}^{k-\ell +1} \left[\frac{m+1}{2} \right]
V_{3\ell + 2m +1}~,
\eeq
with $U_n = u^n - u^{-n}$ and $V_n = v^n - v^{-n}$. $[x]$ stands for the integer
part of $x$. Thus we find the following  $SU(2)_L \times SU(2)_R$ spin contents 
of BPS particle arising from the homology class $2B+kF$;
\beq
\bigoplus_{(j_L, j_R)} N_{2B + kF}^{(j_L, j_R)} \left( j_L, j_R \right) =
\bigoplus_{\ell=1}^k \bigoplus_{m=1}^{k-\ell+1}  \left[\frac{m+1}{2} \right] 
\left( \frac{\ell -1}{2}, \frac{3\ell +2m}{2}\right)~. \label{2inst-spin}
\eeq
For lower values of the winding number $k$, this formula implies
\beqa
k=1 &: & (0, \frac{5}{2}) \CR
k=2 &: & (\frac{1}{2}, 4) \oplus (0, \frac{7}{2}) \oplus (0, \frac{5}{2}) \CR
k=3 &: & (1, \frac{11}{2}) \oplus (\frac{1}{2}, 5)  \oplus (\frac{1}{2}, 4)
\oplus 2 (0, \frac{9}{2}) \oplus (0, \frac{7}{2}) \oplus (0, \frac{5}{2}) \\
k=4 &: &  (\frac{3}{2}, 7) \oplus (1, \frac{13}{2}) \oplus (1, \frac{11}{2})
\oplus 2 (\frac{1}{2}, 6) \oplus (\frac{1}{2}, 5)  \oplus (\frac{1}{2}, 4) 
\oplus 2 (0, \frac{11}{2})  \oplus 2 (0, \frac{9}{2}) \oplus (0, \frac{7}{2}) \oplus (0, \frac{5}{2})
\nonumber
\eeqa
From the formula (\ref{2inst-spin}) we find 
the spin content with the highest left spin is $((k-1)/2, (3k+2)/2)$ for degree $k$. 
Again this is consistent with the moduli space of $D2$ branes. 
The genus of the curves with bi-degree $(2,k)$ is generically
$k-1$ and the moduli space is ${\bf P}^{3k+2}$. 
Thus the generic fiber is $T^{2k-2}$ and the highest left spin is $(k-1)/2$. 
The right spin contents with this highest left spin agrees
with the result from the fact the it is identified with the Lefschetz decomposition
of ${\cal M}_{2B + kF} = {\bf P}^{3k+2}$.
The subleading terms in the spin decomposition are supposed to come from
the degeneration of two cycles to the curve of lower genus. It is an interesting
challenge in mathematics to clarify it 
together with the description of the moduli space of flat line bundles
over degenerate curves. At three instanton level the spin contents 
will get more complicated.
We have checked that the leading spin content for the class $3B + kF$ is
$(0, 7/2), (1,11/2), (2, 15/2), \cdots$, which agree to those expected from the 
geometry of $D$-brane moduli space.

%%%%%%%%%%%%%%%%%%%%%%%%%%%%%%%%%%%%%%%%%%%%%%%%%%%%%%%

\section{Equivariant genus of ${\rm Hilb}^n~{\mathbb C}^2$ : Rank one case}
\setcounter{equation}{0}

In the last section we have seen that the Nekrasov's partition function 
for $SU(2)$ gauge theory carries the information of the refined BPS state counting
in the topological string amplitude on local ${\bf P}^1 \times {\bf P}^1$. It strongly
suggests the existence of two parameter generalization of topological vertex.
In this section we consider $U(1)$ gauge theory with a massive adjoint hypermultiplet
and make an attempt at constructing such a two parameter generalization in terms of the
Macdonald function, which is a two parameter generalization of the Schur function. 
If the Nekrasov's partition function is specialized to the rank one case, it gives the generating 
function of the character of the equivariant cohomology 
of the Hilbert scheme ${\rm Hilb}^n~{\mathbb C}^2$
of $n$ points  on ${\mathbb C}^2$.  In \cite{HIV} it has been argued that five and six dimensional
lifts of the abelian gauge theory with a massive adjoint matter compute
the $\chi_y$ genus and the elliptic genus of ${\rm Hilb}^n~{\mathbb C}^2$,
respectively and that a diagrammatic computation in terms of topological vertex is presented.
Let us briefly review their method in five dimensional theory.

%%%%%%%%%%%%%%%%%%%%%%%%%%%%%%%%%%%%%%%%%%%%%%%
\subsection{$\chi_y$ genus --- Five dimensional theory}

Based on the web-diagram of Fig. 1, we can write down
the partition function in terms of topological vertex $C_{R_1 R_2 R_3}(q)$
that has a parameter $q= e^{- i g_s}$ ($g_s$ is the string coupling constant).
It is given by
\beqa
Z(T, T_m; q) &= & \sum_R e^{-T \cdot |\m^R|} (-1)^{|\m^R|} Z_R (T_m; q)~, \\
Z_R(T_m; q) &=& \sum_{R_m} e^{-T_m \cdot |\m^{R_m}|} (-1)^{|\m^{R_m}|} 
C_{\bullet R_m^t R}(q) C_{\bullet R_m R^t}(q)~,
\eeqa
where the two K\"ahler parameters $T$ and $T_m$ are related to 
the coupling constant $\tau$ of the gauge theory
and the mass of the adjoint hypermultipet as follows;
\beq
Q_\tau := e^{2\pi i \tau} = e^{-T -T_m}~, \qquad Q_m := e^{r m} = e^{-T_m}~.
\eeq
The parameter $r$ is the radius of $S^1$ of the fifth dimension. 
We have identified the representations $R$ and $R_m$ assigned to the (internal) edges
with the partitions $\m^R$ and $\m^{R_m}$, respectively. $|\m^R|$ and $|\m^{R_m}|$
denote the number of boxes of the corresponding Young diagrams.

%%%%%%%%%%%%% Figure one %%%%%%%%%%%%%%%%%%%%
\begin{center}
\begin{pspicture}(-3,-2)(5,4)
\psline(-2,2)(0,2)\psline(0,2)(0,3.5)
\psline(0,2)(2,0)
\psline(2,0)(4,0)\psline(2,0)(2,-1.5)
\psline(-1,1.8)(-1,2.2)\psline(3,-0.2)(3,0.2)
\rput(0.8,0.8){$R_m$}\rput(-1.5,2.3){$R$}\rput(3.5,0.3){$R$}
\rput(0.3,3){$\bullet$}\rput(1.8,-1){$\bullet$}
%\psline[linestyle=dashed,dash=3pt 2pt](1.5,-0.5)(2.5,-0.5)
%\rput(-1.8,1){$R_1$}\rput(-0.4,1){$R_2$}\rput(4.6,1){$R_{N-1}$}\rput(6.2,1){$R_N$}
\end{pspicture}
\end{center}
{\small Figure 1:
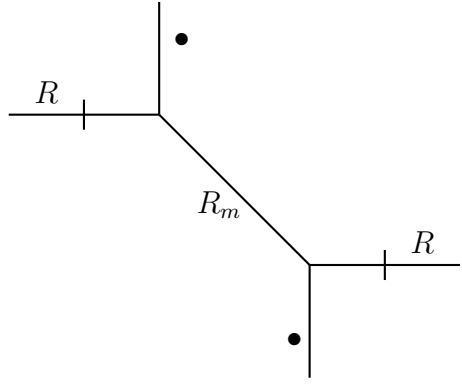
Geometric engineering of five dimensional $U(1)$ theory with adjoint hypermultiplet.
Horizontal external lines are identified to make a $D5$-brane wrapping on a circle,
The (vertical) distance of the horizontal lines is identified with 
the mass of the adjoint matter
}
%%%%%%%%%%%%%%%%%%%%%%%%%%%%%%%%%%%%%%%%%
\vspace{10mm}

Using the expression of the topological vertex $C_{R_1 R_2 R_3}(q)$
in terms of the Schur function $s_R(x)$ and the Cauchy formula;
\beq
\sum_R s_R(x) s_{R^t} (y) = \prod_{i,j} (1 + x_i y_j)~,
\eeq
we obtain
\beqa
Z_R(T_m, q) &=& s_R(q^\rho) s_{R^t} (q^\rho) \prod_{i,j} (1 - Q_m q^{\mu^R_i + \mu_j^{R^t} -i-j +1}) \CR
&=& s_R (q^\rho) s_{R^t} (q^\rho) \prod_{k \geq 1} (1-Q_m q^k)^k
\prod_{(i,j) \in \mu^R} (1 - Q_m q^{h(i,j)}) (1 - Q_m q^{-h(i,j)})~,
\eeqa
where $h(i,j)$ is the hook length at the box $(i,j)$ in the Young diagram of $\mu^R$.
Substituting the specialization of the Schur function;
\beq
s_R(q^\rho) s_{R^t}(q^\rho) = \frac{ q^{\sum_{(i,j)\in \m^R} h(i,j)}}
{\prod_{(i,j)\in \m^R} (1 - q^{h(i,j)})^2}~,
\eeq
we finally find
\beqa
Z &=& \prod_{k \geq 1} (1-Q_m q^k)^k \sum_R Q^{|\m^R|}
\prod_{(i,j)\in \m^R} \frac{(1-Q_m q^{h(i,j)})(1-Q_m q^{-h(i,j)})}
{(1 - q^{h(i,j)})(1- q^{-h(i,j)})} \\
&=& \prod_{k \geq 1} (1-Q_m q^k)^k \sum_R Q_\tau^{|\m^R|}
\prod_{(i,j)\in \m^R} \frac{(1-Q_m q^{h(i,j)})(1-Q_m^{-1} q^{h(i,j)})}
{(1 - q^{h(i,j)})^2}~, \label{HIV}
\eeqa
with $Q_\tau = Q Q_m$.

The generating function of the equivariant $\chi_y$ genera of 
${\rm Hilb}^n~{\mathbb C}^2$ is given by \cite{LLZ};
\beq
\sum_{n=0}^\infty Q^n \chi_y \left( {\rm Hilb}^n~{\mathbb C}^2 \right)(t,q)
= \sum_\m Q^{|\m|} \prod_{s\in \m} 
\frac{(1-y~t^{-\ell(s)}q^{-a(s)-1})(1-y~t^{\ell(s)+1}q^{a(s)})}
{(1- t^{-\ell(s)}q^{-a(s)-1})(1- t^{\ell(s)+1}q^{a(s)})}~,   \label{chiy}
\eeq
where $t=t_1$ and $q^{-1} = t_2$ are equivariant parameters of 
the toric action. When $q=t$ this generating function agrees with 
the instanton part of the partition function (\ref{HIV})\footnote{The first
factor in (\ref{HIV}) is the perturbative contribution.},
if we identify $y=Q_m$. In the decoupling limit
of the adjoint matter $m \to \infty$,  the expression is reduced to
the generating function of the equivariant Euler character $\chi_0$;
\beqa
\sum_{n=0}^\infty Q^n \chi_0 \left( {\rm Hilb}^n~{\mathbb C}^2 \right)(t,q)
&=& \sum_\m Q^{|\m|} \prod_{s\in \m} 
\frac{1}{(1- t^{-\ell(s)}q^{-a(s)-1})(1- t^{\ell(s)+1}q^{a(s)})}~,  \CR
&=& \exp \left( \sum_{n=1}^\infty \frac{Q^n}{n(1-t^n)(1-q^{-n})} \right)~.
 \label{chi0}
\eeqa
We note that this is identified as the partition function of the abelian gauge theory
that is given by the formula (\ref{5Dlift}) if we (formally) extend it to $U(1)$ case .

%%%%%%%%%%%%%%%%%%%%%%%%%%%%%%%%%%%%%%%%%%%%%%%%%%%%
\subsection{Two parameter generalization}

We will show that the full generating function (\ref{chiy}) 
is obtained by topological vertex computation, if we use a two
parameter generalization of topological vertex $C_{R_1 R_2 R_3}(q,t)$
in terms of the Macdonald function $P_R(x;q,t)$. 
We introduce the refined topological vertex
\beq
C_{R_1 R_2 R_3}(q,t) := t^{-n(R_3^t)} q^{n(R_3)}
 {\widetilde P}_{R_2}(t^{-\rho}; q,t) \sum_R \iota
 {\widetilde P}_{R_1/R}(q^{+\m^{R_2}} t^{+\rho}; q,t)
 {\widetilde P}_{R_3^t/R}(q^{-\m^{R_2}} t^{-\rho}; q,t)~, \label{rtv}
\eeq
and its conjugate
\beq
C^{R_1 R_2 R_3}(q,t) := C_{R_1 R_2 R_3}(t,q)~,
%&:=& t^{-n(R_3^t)} q^{n(R_3)} {\widetilde P}_{R_2}(\alpha q^{-\rho}; t,q) \sum_R \iota
%{\widetilde P}_{R_1/R} ( \alpha t^{+\m^{R_2}} q^{+\rho}; t,q)
%{\widetilde P}_{R_3^t/R}( \alpha t^{-\m^{R_2}} q^{-\rho}; t,q)~, \nonumber
\eeq
where we have exchanged $q$ and $t$ to define the conjugate vertex. 
We need two types of topological vertices related by the exchange of $q$ and $t$ and
will use lower and upper indices to distinguish them.
We find it convenient to use a renormalized (skew) Macdonald function
$\widetilde P_{R/S}(x; q,t)$ defined as follows; Let us introduce the normalization factor
\beq
f_R^2(q,t) =  b_R(q,t) := 
\prod_{s \in \m^R} \frac{1- q^{a(s)} t^{\ell(s) +1} } {1- q^{a(s)+1} t^{\ell(s)}}~.
\eeq
We note the relation
\beq
f_R(q,t) f_{R^t} (t, q) = 1~.
\eeq
Then the renormalized skew Macdonald function is given by
\beq
{\widetilde P}_{R/S}(x ; q, t) := f_R(q,t) f_S^{-1}(q,t) P_{R/S} (x;q,t)~.
\eeq
In (\ref{rtv}) the notation $\iota$ on the second Macdonald function means 
the involution on the symmetric functions defined by $\iota(p_n)= - p_n$ for
the power sum function $p_n(x) = \sum_i x_i^n$.
Note that the set of the power sum functions $\{ p_n(x) \}_{n\geq 0}$ 
generates the ring of symmetric functions. 
Finally $q^{\m} t^{\rho}$ stands for the specialization
with $x_i = q^{\m_i} t^{-i + 1/2}$.  
One may notice that the first index $R_1$ is distinguished from 
other indices and we cannot prove the cyclic symmetry of our topological vertex.
To construct a refined vertex with nice symmetry is an open problem.

When $q=t$, the (renormalized) Macdonald function is reduced to
the Schur function $s_R(x)$. Hence, the refined vertex (\ref{rtv}) becomes
\beqa
C_{R_1 R_2 R_3}(q) &=& C^{R_1 R_2 R_3}(q) \CR
&=& q^{n(R_3)-n(R_3^t)}
 s_{R_2}(q^{-\rho}) \sum_R \iota s_{R_1/R}(q^{+\m^{R_2} +\rho} )
 s_{R_3^t/R}(q^{-\m^{R_2}-\rho})~. \label{TV}
\eeqa
Using the formula of analytic continuation for the specialization of the Schur function;
\beq
s_{R/Q} (q^{\m^t + \rho}) = \iota s_{R/Q} (q^{-\m -\rho})~,
\eeq
and the relation $\kappa (\m) = 2(n(\m^t) - n(\m))$,
we see that (\ref{TV}) agrees with the topological vertex in terms of
the skew Schur functions \cite{ORV};
\beq
C_{R_1 R_2 R_3}(q) = q^{- \kappa(R_3)/2}
 s_{R_2}(q^{-\rho}) \sum_R s_{R_1/R}(q^{- \m^{R_2^t} - \rho} )
 s_{R_3^t/R}(q^{-\m^{R_2}-\rho})~. 
\eeq

With the refined topological vertex (\ref{rtv}) 
and the same web diagram as before, we have the topological partition function
\beqa
Z(Q, Q_m ; q, t) &=& \sum_{R, R_m} (-Q)^{|\m^R|} (-Q_m)^{|\m^{R_m}|}
C_{ \bullet R R_m}(q,t) C^{\bullet R^t R_m^t }(q,t) \CR
&=& \sum_R  (- Q)^{|\m^R|} P_{R} (t^{-\rho};q,t) P_{R^t} (q^{-\rho};t,q) \CR
& &~~~\sum_{R_m} (- Q_m)^{|\m^{R_m}|}
P_{R_m^t} (q^{-\m^R} t^{-\rho};q,t) P_{R_m} (t^{-\m^{R^t}} q^{-\rho};t,q)~,
\eeqa
where we have used ${\widetilde P}_\m (x;q,t){\widetilde P}_{\m^t} (y;t,q)
= P_\m (x;q,t) P_{\m^t} (y;t,q)$. Since we can take the first index to be trivial,
we can avoid the problem of asymmetry mentioned above in this case. 
By the specialization formula in Appendix B.3
we have
\beqa
P_{\lam}(t^{-\rho} ; q,t) &=& \frac{t^{|\lam|/2 + n(\lam)}}
{\prod_{s \in \lam} ( 1 - q^{a(s)} t^{\ell(s) + 1}) }~, \CR
P_{\lam^t}(q^{-\rho} ; t,q) &=& \frac{q^{|\lam|/2 + n(\lam^t)}}
{\prod_{s \in \lam} ( 1 - t^{\ell(s)} q^{a(s) + 1}) }~, \CR
&=& \frac{(-1)^{|\lam|} q^{|\lam|/2 + n(\lam^t)}}
{ t^{n(\lam)} q^{n(\lam^t) + |\lam|}\prod_{s \in \lam} ( 1 - t^{-\ell(s)} q^{-a(s) - 1}) }~.
\eeqa
Hence we obtain
\beqa
& & \sum_R (- Q)^{|\m^R|} P_{R} ( t^{-\rho} ; q,t) P_{R^t}  ( q^{-\rho} ; t,q)  \CR
&=& \sum_R  Q^{|\m^R|} \left( \frac{t}{q} \right)^{|\m^R|/2} 
\prod_{s \in \m^R} \frac{1}{( 1 - q^{a(s)} t^{\ell(s) + 1}) ( 1 - t^{-\ell(s)} q^{-a(s) - 1}) }~. \label{sum}
\eeqa
We next invoke the Cauchy formula to obtain
\beq
\sum_{R_m}  (- Q_m)^{|\m^R|} P_{R_m^t} (q^{-\m^R} t^{-\rho} ; q,t) 
P_{R_m}  (t^{-\m^{R^t}} q^{-\rho} ; t,q) 
= \prod_{i,j \geq 1} \left( 1 - Q_m q^{-\m_i^R + j - \frac{1}{2} } 
t^{-\m_j^{R^t} + i - \frac{1}{2}} \right)~.
\eeq
Combining two propositions in Appendix A.2, we can derive
\beq
\sum_{s \in \m} q^{a_\lam(s)} t^{\ell_\m(s) +1} 
+ \sum_{s \in \lam} q^{-a_\m(s) -1} t^{- \ell_\lam(s)}
= \sum_{i,j \geq 1} \left( q^{-\m_i} t^{-\lam_j^t} -1 \right) t^i q^{j-1}~.
\label{convert}
\eeq
The equation (\ref{convert}) with $\m = \lam$ implies
\beqa
& &\prod_{i,j \geq 1} \left( 1 -  Q_m q^{-\m_i^R + j - \frac{1}{2} } 
t^{-\m_j^{R^t} + i - \frac{1}{2}} \right)  \\
&=& \prod_{i,j \geq 1} \left( 1 -  Q_m q^{j - \frac{1}{2}} t^{i - \frac{1}{2}} \right)
\cdot \prod_{s \in \m^R} \left( 1 - Q_m \left( \frac{q}{t} \right)^{1/2}
q^{a(s)} t^{\ell(s) + 1} \right) \left( 1 - Q_m \left( \frac{q}{t} \right)^{1/2}
t^{-\ell(s)} q^{-a(s) - 1} \right)~. \nonumber
\eeqa
Combining all these results we have the generating function (\ref{chiy}) 
of the $\chi_y$ genus with the identification $y = Q_m \left( \frac{q}{t} \right)^{1/2}$.

%%%%%%%%%%%%%%%%%%%%%%%%%%%%%%%%%%%%%%%%%%%%%%%%%%%%
\subsection{Elliptic genus --- Six dimensional theory}

According to the web-diagram (Fig. 2) of the six dimensional theory, 
the topological partition function is 
\beqa
Z(Q, Q_v, Q_m ; q, t) &=& \sum_{R, R_v, R_m} (-Q)^{|\m^R|}  (-Q_v)^{|\m^{R_v}|} (-Q_m)^{|\m^{R_m}|}
C_{R_v R R_m}(q,t) C^{R_v^t  R^t R_m^t }(q,t) \CR
&=& \sum_R  (- Q)^{|\m^R|} P_{R} (t^{-\rho};q,t) P_{R^t} (q^{-\rho};t,q) Z_R(Q_v, Q_m ; q, t)~,
\eeqa
where
\beqa
& & Z_R (Q_v, Q_m ; q, t) = \sum_{R_v, R_m} (- Q_v)^{|\m^{R_v}|} (- Q_m)^{|\m^{R_m}|}
 \sum_{T_1, T_2} \iota {\widetilde P}_{R_v/T_1}(q^{\m^R} t^\rho ; q, t) 
 {\widetilde P}_{R_m^t/T_1}(q^{-\m^R} t^{-\rho} ; q, t)   \CR
& &~~~ \times
\iota {\widetilde P}_{R_v^t/T_2}( t^{\m^{R^t}} q^\rho ; t, q) 
 {\widetilde P}_{R_m/T_2}( t^{-\m^{R^t}} q^{-\rho} ; t, q) ~.
\eeqa
We note that the prefactor $t^{-n(R_3^t)} q^{n(R_3)}$ in our definition (\ref{rtv}) of
the refined topological vertex plays no role here. We expect this factor is important
when we consider the web diagram with non-trivial framing.

%%%%%%%%%%%%% Figure two %%%%%%%%%%%%%%%%%%%%
\begin{center}
\begin{pspicture}(-3,-2)(5,4)
\psline(-2,2)(0,2)\psline(0,2)(0,3.5)
\psline(0,2)(2,0)
\psline(2,0)(4,0)\psline(2,0)(2,-1.5)
\psline(-1,1.8)(-1,2.2)\psline(3,-0.2)(3,0.2)
\psline(-0.2,3)(0.2,3)\psline(-0.2,3.2)(0.2,3.2)
\psline(1.8,-1)(2.2,-1)\psline(1.8,-1.2)(2.2,-1.2)
\rput(0.8,0.8){$R_m$}\rput(-1.5,2.3){$R$}\rput(3.5,0.3){$R$}
\rput(0.5,3){$R_v$}\rput(1.5,-1){$R_v$}
%\psline[linestyle=dashed,dash=3pt 2pt](1.5,-0.5)(2.5,-0.5)
%\rput(-1.8,1){$R_1$}\rput(-0.4,1){$R_2$}\rput(4.6,1){$R_{N-1}$}\rput(6.2,1){$R_N$}
\end{pspicture}
\end{center}
{\small Figure 2:
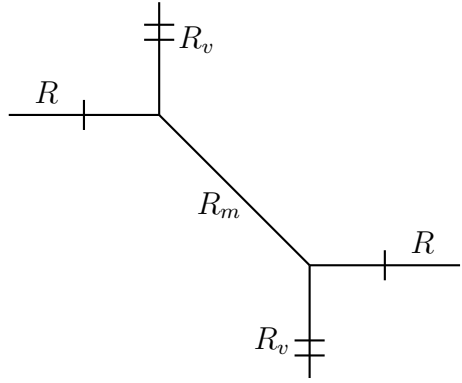
Geometric engineering of six dimensional $U(1)$ theory with adjoint hypermultiplet.
In addition to the horizontal external lines the vertical external lines ($NS~5$-branes)
are identified.
}
%%%%%%%%%%%%%%%%%%%%%%%%%%%%%%%%%%%%%%%%%
\vspace{10mm}

In the following computation of $Z_R$, we will repeatedly employ the Cauchy formula for the skew
Macdonald function;
\beqa
& & \sum_R {\widetilde P}_{R/R_1} (x;q,t) {\widetilde P}_{R/R_2} (y;q,t) 
= \Pi(x,y ; q,t) \sum_S {\widetilde P}_{R_2/S} (x;q,t) {\widetilde P}_{R_1/S} (y;q,t)~, \\
& & \sum_R {\widetilde P}_{R/R_1} (x;q,t) {\widetilde P}_{R^t/R_2^t} (y;t,q) 
= \Pi_0 (x,y) \sum_S {\widetilde P}_{R_2/S} (x;q,t) {\widetilde P}_{R_1^t/S^t} (y;t,q)~,
\eeqa
where
\beqa
\Pi (x,y ; q,t) := \prod_{i,j \geq 1} \frac{(tx_i y_j ; q)_\infty}{(x_i y_j ; q)_\infty} 
= \exp \left( \sum_{n=1}^\infty \frac{1}{n} \frac{1-t^n}{1-q^n} p_n(x) p_n(y) \right)~, \label{pi1}  \\
\Pi_0 (x,y) := \prod_{i,j \geq 1} ( 1 + x_i y_j )  
= \exp \left( \sum_{n=1}^\infty \frac{(-1)^{n-1}}{n} p_n(x) p_n(y) \right)~.  \label{pi2} 
\eeqa
Recall the definition of the involution $\iota$; $\iota(p_n) = - p_n$. 
Looking at (\ref{pi1}) and (\ref{pi2}), we see the following Cauchy formula
with involution;
\beqa
& & \sum_R \iota{\widetilde P}_{R/R_1} (x;q,t) {\widetilde P}_{R/R_2} (y;q,t)
= \Pi(x,y ; q,t)^{-1}  \sum_S  \iota{\widetilde P}_{R_2/S} (x;q,t) {\widetilde P}_{R_1/S} (y;q,t)~, \\
& & \sum_R \iota{\widetilde P}_{R/R_1} (x;q,t) \iota{\widetilde P}_{R^t/R_2^t} (y;t,q) 
= \Pi_0 (x,y) \sum_S \iota{\widetilde P}_{R_2/S} (x;q,t) \iota{\widetilde P}_{R_1^t/S^t} (y;t,q)~.
\eeqa
Now we are ready for computing $Z_R$. By the Cauchy formula, we have
\beqa
& & Z_R(Q_v, Q_m : q,t) = \prod_{i,j \geq 1} \left( 1 - Q_v q^{\m_i^R - j + \frac{1}{2} } 
t^{\m_j^{R^t} - i + \frac{1}{2}} \right) \left( 1 -  Q_m q^{-\m_i^R + j - \frac{1}{2} } 
t^{-\m_j^{R^t} + i - \frac{1}{2}} \right) \CR
& & \sum_{S_1 S_2} (-Q_v)^{|\m^{S_1}|}  (-Q_m)^{|\m^{S_2}|} 
\sum_{T_1 T_2} \iota {\widetilde P}_{T_2^t/S_1} (Q_v q^{\m^R} t^\rho ; q,t) \CR
& & \times \iota {\widetilde P}_{T_1^t/S_1^t} ( Q_v  t^{\m^{R^t}} q^\rho ; t,q) 
{\widetilde P}_{T_2^t/S_2} ( Q_m q^{-\m^{R}} t^{-\rho} ; q,t) 
{\widetilde P}_{T_1^t/S_2^t} ( Q_m t^{-\m^{R^t}} q^{-\rho} ; t,q) \CR
&=& \prod_{i,j \geq 1} \left( 1 - Q_v q^{\m_i^R - j + \frac{1}{2} } 
t^{\m_j^{R^t} - i + \frac{1}{2}} \right) \left( 1 - Q_m q^{-\m_i^R + j  - \frac{1}{2}} 
t^{-\m_j^{R^t} + i  - \frac{1}{2}} \right) \CR
& & \Pi^{-1} \left( Q_v q^{\m^R} t^\rho,  Q_m q^{-\m^{R}} t^{-\rho}; q,t \right) 
 \Pi^{-1} \left( Q_v  t^{\m^{R^t}} q^\rho, Q_m t^{-\m^{R^t}} q^{-\rho} ; t,q \right) \CR
& & \sum_{S_1 S_2} (- Q_v)^{|\m^{S_1}|}  ( - Q_m)^{|\m^{S_2}|} 
\sum_{T_3 T_4} 
\iota {\widetilde P}_{S_2/T_3} (Q_v q^{\m^R} t^\rho ; q,t) \CR
& & \times {\widetilde P}_{S_1/T_3} (Q_m q^{-\m^{R}} t^{-\rho} ; q,t) 
\iota {\widetilde P}_{S_2^t/T_4} ( Q_v  t^{\m^{R^t}} q^\rho ; t,q) 
{\widetilde P}_{S_1^t/T_4} ( Q_m t^{-\m^{R^t}} q^{-\rho} ; t,q)~.
\eeqa
Our rule of analytic continuation explained in Appendix A.2 implies
\beqa
\Pi (q^\lam t^\rho, q^{-\m} t^{-\rho} ; q, t)
&=& \Pi_0 ( - t^{-\lam^t} q^{-\rho - \frac{1}{2}}, q^{-\m} t^{- \rho + \frac{1}{2}} ) \CR
&=& \Pi_0 ( q^{\lam} t^{\rho + \frac{1}{2}}, - t^{\m^t} q^{\rho- \frac{1}{2}} )~.
\eeqa
Hence we obtain
\beqa
Z_R
&=& \prod_{i,j \geq 1} \frac{ \left( 1 - Q_v q^{\m_i^R - j + \frac{1}{2} } 
t^{\m_j^{R^t} - i  + \frac{1}{2}} \right) \left( 1 - Q_m q^{-\m_i^R + j  - \frac{1}{2}} 
t^{-\m_j^{R^t} + i  - \frac{1}{2}} \right)}
{ \left( 1 - Q_v Q_m q^{\m_i^R - j}  t^{\m_j^{R^t} - i + 1} \right) 
\left( 1 - Q_v Q_m q^{-\m_i^R + j} t^{-\m_j^{R^t} + i -1} \right) } \CR
& & \sum_{S_1 S_2} (-Q_v)^{|\m^{S_1}|}  (-Q_m)^{|\m^{S_2}|} 
\sum_{T_3 T_4} \iota {\widetilde P}_{S_2/T_3} (Q_v  q^{\m^R} t^\rho ; q,t) \CR
& & \times {\widetilde P}_{S_1/T_3} (Q_m q^{-\m^{R}} t^{-\rho} ; q,t) 
\iota {\widetilde P}_{S_2^t/T_4} ( Q_v t^{\m^{R^t}} q^\rho ; t,q) 
{\widetilde P}_{S_1^t/T_4} ( Q_m t^{-\m^{R^t}} q^{-\rho} ; t,q)~.
\eeqa
Here we notice that the sum over the representations is the same as before except
that the arguments of $\iota {\widetilde P}$ and ${\widetilde P}$ are multiplied 
by $Q_v$ and $Q_m$, respectively. Thus we can repeat
the same procedure of using the Cauchy formula. By iteration we finally obtain
\beq
Z_R = \prod_{k=1}^\infty  \prod_{i,j \geq 1} 
\frac{ \left( 1 - Q_v^k Q_m^{k-1} q^{\m_i^R - j +  \frac{1}{2}} t^{\m_j^{R^t} - i + \frac{1}{2} } \right) 
\left( 1 - Q_v^{k-1} Q_m^k q^{-\m_i^R + j  - \frac{1}{2}} t^{-\m_j^{R^t} + i  - \frac{1}{2}} \right)} 
{ \left( 1 - Q_v^k Q_m^k q^{\m_i^R - j} t^{\m_j^{R^t} - i + 1} \right)
 \left( 1 -  Q_v^k Q_m^k q^{-\m_i^R + j} t^{-\m_j^{R^t} + i-1} \right) }~.
\eeq

The formula (\ref{convert}) with $\lam =\m$
\beqa
\sum_{s \in \m} \left(  q^{a(s)} t^{\ell(s) + 1} + q^{-a(s) -1} t^{-\ell(s)} \right)
&=& \sum_{i,j \geq 1} \left( q^{- \m_i} t^{-\m_j^t} -1  \right) t^i q^{j-1}~, \\
\sum_{s \in \m} \left(  q^{-a(s)} t^{-\ell(s) - 1} + q^{a(s) +1} t^{\ell(s)} \right)
&=& \sum_{i,j \geq 1} \left( q^{ \m_i} t^{ \m_j^t} -1  \right) t^{-i} q^{-j +1}~,
\eeqa
allows us to convert the infinite product over two indices $i,j$ into the finite
product over the boxes of the Young diagram;
\beqa
Z_R &=& Z_{perturb} \cdot \prod_{k=1}^\infty \prod_{s \in \m^R} 
\frac {\left( 1 - Q_v^k Q_m^{k-1} \left( \frac{t}{q} \right)^{\frac{1}{2}} q^{-a(s)} t^{-\ell(s) - 1} \right) 
\left( 1 - Q_v^k Q_m^{k-1}  \left( \frac{t}{q} \right)^{\frac{1}{2}} q^{a(s)+1} t^{\ell(s)} \right)}
{\left( 1 - Q_v^k Q_m^k q^{-a(s)-1} t^{- \ell(s)} \right) 
\left( 1 - Q_v^k Q_m^k q^{a(s)} t^{\ell(s) + 1} \right) } \CR
& &~~~ \times \frac{
\left( 1 - Q_v^{k-1} Q_m^k \left( \frac{q}{t} \right)^{\frac{1}{2}} q^{a(s)} t^{\ell(s)+1}  \right)
\left( 1 - Q_v^{k-1} Q_m^k \left( \frac{q}{t} \right)^{\frac{1}{2}} q^{-a(s) - 1} t^{-\ell(s)}  \right) }
{\left( 1 - Q_v^k Q_m^k  q^{ a(s) + 1} t^{\ell(s)} \right)
\left( 1 - Q_v^k Q_m^k q^{- a(s)} t^{- \ell(s) - 1} \right)  }~,
\eeqa
where
\beq
Z_{perturb} =
\prod_{k=1}^\infty  \prod_{i,j \geq 1} \frac {\left( 1 - Q_v^k Q_m^{k-1}
 q^{-j + \frac{1}{2}} t^{- i + \frac{1}{2}} \right)
\left( 1 - Q_v^{k-1} Q_m^k q^{j - \frac{1}{2}} t^{i - \frac{1}{2}} \right)}
{\left( 1 - Q_v^k Q_m^k q^{- j} t^{- i+1} \right)
\left( 1 - Q_v^k Q_m^k q^{j} t^{i -1} \right)}~.
\eeq
and we have neglected the factor $\prod_{k=1}^\infty (1-Q_v^k Q_m^k)^{-1}$ that is independent of $q$ and $t$.
Combined with the contribution of (\ref{sum}), the instanton part of $Z$ is identified as follows;
\beqa
Z(Q, y, p ; q,t) &=&  \sum_R  Q^{|\m^R|} \left( \frac{t}{q} \right)^{|\m^R|/2} 
\prod_{k=1}^\infty  \prod_{s \in \m^R} 
\frac {\left( 1 -  p^k y^{-1} q^{-a(s)} t^{-\ell(s) - 1} \right) 
\left( 1 - p^k y^{-1}  q^{a(s)+1} t^{\ell(s)} \right)}
{\left( 1 - p^{k-1}  q^{-a(s)-1} t^{- \ell(s)} \right) 
\left( 1 - p^{k-1} q^{a(s)} t^{\ell(s) + 1} \right) } \CR
& &~~~ \times \frac{
\left( 1 - p^{k-1} y q^{a(s)} t^{\ell(s)+1}  \right)
\left( 1 - p^{k-1} y  q^{-a(s) - 1} t^{-\ell(s)}  \right) }
{\left( 1 - p^{k} q^{ a(s) + 1}  t^{\ell(s)} \right)
\left( 1 - p^{k}  q^{- a(s)} t^{- \ell(s) - 1} \right)  }~,
\eeqa
where $p:=Q_v Q_m$ and $y:= Q_m \cdot \left( \frac{q}{t} \right)^{\frac{1}{2}}$.
We can see the final result agrees to the generating function of 
the elliptic genus of ${\rm Hilb}^n~{\mathbb C}^2$ given in \cite{LLZ}
with $t = t_1^{-1}$ and $q=t_2$.

%%%%%%%%%%%%%%%%%%%%%%%%%%%%%%%%%%%%%%%%%%%%%%%%%%%%%%%%%%%%%%

\vskip1cm

We would like to thank T.~Eguchi, S.~Hosono, Y.~Konishi, H.~Nakajima, 
N.~Nekrasov, A~Okounkov, Y.~Tachikawa and Jian Zhou for helpful discussions.
Research of H.A. and H.K. is supported in part 
by the Grant-in Aid for Scientific Research (No. 13135212 and No.14570073) from 
Japan Ministry of Education, Culture and Sports. 

%%%%%%%%%%%%%%%%%%%%%%%%%%%%%%%%%%%%%%%%%%%%%%%%%%

%%%%%%%%%%%%%%%%%%% letter %%%%%%%%%%%%%%%%%%%%%%%%%%%%%%%%%%%
%\def\e{\epsilon}
%\def\i{\iota}
\newcommand\la{\lambda}
%\def\r{\rho}
%\def\s{\sigma}
%\def\bC{\mathbb C}%{{\bf C}}
%\def\bZ{{\bf Z}}    
%%%%%%%%%%%%%%%%%% equation %%%%%%%%%%%%%%%%%%%%%%%%%%%%%%%%%%%%%%%%%%
\newcommand\be{\begin{equation}}
\newcommand\ee{\end{equation}}
\newcommand\ba{\begin{eqnarray}}
\newcommand\ea{\end{eqnarray}}

\newcommand{\Smat}[1]{\left[\matrix{#1}\right]}

\newcommand\ha{\frac{1}{2}}
%\renewcommand\->{\rightarrow}
%\newcommand{\Exp}[1]{\exp\left\{#1\right\}}
%%%%%%%%%%%%%%%%% def only for this article %%%%%%%%%%%%%%%%%%%%%%
%\def\tF{\widetilde F}
%\def\tQ{\widetilde Q}
\newcommand\tP{\widetilde P}

\setcounter{section}{0}

%%%%%%%%%%%%%%%%%%%%%%%%%%%%%%%%%%%%%%%%%%%%%%%%%%%%%%%%%%%%

\section*{Appendix A : Formula for Partition}
\renewcommand{\theequation}{A.\arabic{equation}}\setcounter{equation}{0}
\renewcommand{\thesubsection}{A.\arabic{subsection}}\setcounter{subsection}{0}

%%%%%%%%%%%%%%%%%%%%%%%%%%%%%%%%%%%%%%%%%%%%%%%%%%%%%%%%%%%%

%%%%%%%%%%%%%%%%%%%%%%%%%%%%%%%%%%%%%%%%%%%%%%%%%%

\subsection{ Arm-length, leg length and related quantities}

%%%%%%%%%%%%%%%%%%%%%%%%%%%%%%%%%%%%%%%%%%%%%%%%%%

For each square  $s=(i,j)$ in the Young diagram of 
a partition $\{ \lam_i \}$, we define
\beq
a_\lam(s) := \lam_i - j, \quad \ell_\lam(s) := \lam_j^\vee - i, 
\quad a'(s) := j -1, \quad \ell'(s) := i -1~,
\eeq
where $\lam_j^\vee$ denotes the conjugate (dual) diagram. 
They are called arm-length, leg-length, arm-colength and leg-colength,
respectively.
The hook length $h_\lam(s)$ and the content $c(s)$ at $s$ are given by
\beq
h_\lam(s) = a_\lam(s) + \ell_\lam(s) + 1~, \quad c(s) = a'(s) - \ell'(s)~.
\eeq
We also need the following integer
\beq
n(\lam)  := \sum_{s \in \lam} \ell'(s) = \sum_{i=1}^\infty (i-1) \lam_i
= \frac{1}{2} \sum_{i=1}^\infty \lam_i^\vee ( \lam_i^\vee -1) 
= \sum_{s \in \lam} \ell_\lam (s)~.
\eeq
Similarly we have
\beq
n(\lam^\vee)  := \sum_{s \in \lam} a'(s) = \sum_{s \in \lam} a_\lam (s)~.
\eeq
They are related to the integer $\k(\lam)$ as follows;
\beq
\k(\lam) := 2 \sum_{s \in \lam} (j-i) = 2(n(\lam^\vee) - n(\lam))
= |\lam| + \sum_{i=1}^\infty \lam_i (\lam_i -2i)~.
\eeq

%%%%%%%%%%%%%%%%%%%%%%%%%%%%%%%%%%%%%%%%%%%%%%%%%%%%%%%%

\subsection{Combinatorial identities} 

%%%%%%%%%%%%%%%%%%%%%%%%%%%%%%%%%%%%%%%%%%%%%%%%%%%%%%%%

Next we will prove two useful propositions.
First, we have
\\
%%%%%%%%%%%%%% proposition %%%%%%%%%%%%%%
{\bf Lemma  1.}
{\it For all integers $N\geq\ell(\la )$}
\be
(1-q)\sum_{(i,j)\in\la } q^{j-1} t^{-i+1} 
=
\sum_{i=1}^N \left(1-q^{\la_i}\right) t^{-i+1},
\label{eq:partitionFormulaI}
\ee
\ba
(1-q)\sum_{(i,j)\in\mu } q^{\la_i-j} t^{\mu^\vee_j-i+1} 
&=&
(t-1)\sum_{1\leq i<j\leq N+1} \left(q^{\la_i-\mu_j}-1 \right)t^{j-i} +
t\sum_{i=1}^N \left(q^{\la_i-\mu_i}- 1\right)
\cr
&-& 
\sum_{i=1}^N \left(q^{\la_i} - 1 \right) t^{N-i+2}.
%(t-1)\sum_{1\leq i<j\leq N+1} q^{\la_i} \left(q^{-\mu_j}-1 \right)t^{j-i} +
%t\sum_{i=1}^N q^{\la_i} \left(q^{-\mu_i}- 1\right),
\label{eq:partitionFormulaII}
\ea
%%%%%%%%%%%%%% end of proposition %%%%%%%%%%%%%%
%
The former in this lemma is nothing but
$\sum_{j=1}^{\la} q^{j-1} = {(1-q^{\la })/(1-q)}$.
The proof of the latter %(\ref{eq:partitionFormulaII}) 
is similar to that in \cite{Mac}.

%%%%%%%%%%%%%% end of proof %%%%%%%%%%%%%%

When $|t|<1$, the last term of (\ref{eq:partitionFormulaII}) vanishes 
as $N$ tends to infinity. 
Using this formula and that replacing 
$q$, $t$ and $\la $ with
$1/q$, $1/t$ and $\mu $, respectively,
we obtain 
\\
%%%%%%%%%%%%%% proposition %%%%%%%%%%%%%%
{\bf Proposition 1.}
\beq
\sum_{s \in \m} q^{a_\lam(s)} t^{\ell_\m (s) +1} + 
\sum_{s \in \lam} q^{-a_\m(s) - 1} t^{- \ell_\lam (s)} 
=\frac{t-1}{1-q} \sum_{i,j = 1}^\infty 
\left( q^{\lam_i - \m_j} - 1 \right) t^{j-i}~.
\eeq
%%%%%%%%%%%%%% end of proposition %%%%%%%%%%%%%%

By the exchange $(\lam, \m) \to (\lam^\vee, \m^\vee)$ and $(q,t) \to (t,q)$,
we obtain the transposed version;
\beq
\sum_{s \in \m} q^{a_\m(s) +1} t^{\ell_\lam (s)} + 
\sum_{s \in \lam} q^{-a_\lam(s)} t^{- \ell_\m (s) -1} 
=\frac{q-1}{1-t} \sum_{i,j = 1}^\infty  
\left( t^{\lam_i^\vee - \m_j^\vee} - 1 \right) q^{j-i}~.
\eeq

In this formula we are aware of the problem of the domain of the convergence
of the geometric series in $t$.
We understand that the geometric series is
computed in an appropriate domain in the complex $t$-plane and then
analytically continued to the whole plane as  rational function with 
a pole at $t=1$ (and $q=1$). 

Next, similar to (\ref{eq:partitionFormulaI}),  we have
\\
%%%%%%%%%%%%%% proposition %%%%%%%%%%%%%%
{\bf Lemma  2.}
{\it For all integers $N\geq\ell(\la )$ and $M\geq\la_1$,}
\be
\left(t^{\ha}-t^{-\ha}\right)\sum_{i=1}^N q^{\la_i}t^{\ha -i}
+
\left(q^{\ha}-q^{-\ha}\right)\sum_{i=1}^M t^{-\la^\vee_i}q^{i-\ha }
= 
q^M - t^{-N}.
\ee
%%%%%%%%%%%%%% end of proposition %%%%%%%%%%%%%%
Where $\ell(\la )$ is the length of the partition $\la$,
which is the non-zero numbers of parts $\la_i$'s. 

When $|q|$, $|t^{-1}|<1$,  the right hand side of (\ref{eq:AC}) vanishes 
as $N$ and $M$ tends to infinity. 
Hence we obtain
\\
%%%%%%%%%%%%%% proposition %%%%%%%%%%%%%%
{\bf Proposition 2.}
\beq
(t-1) \sum_{i=1}^\infty q^{\lam_i} t^{-i} 
= (q^{-1} -1) \sum_{i=1}^\infty t^{-\lam_i^\vee} q^{i}~.
\label{eq:AC}
\eeq
%%%%%%%%%%%%%% end of proposition %%%%%%%%%%%%%%

Taking $(q,t) \to (q^{-1}, t^{-1})$ we have the analytically continued version;
\beq
(t^{-1}-1) \sum_{i=1}^\infty q^{- \lam_i} t^{i} 
= (q -1) \sum_{i=1}^\infty t^{\lam_i^\vee} q^{-i}~.
\eeq

When $q=t$ these propositions become
\beqa
\sum_{s \in \m} q^{a_\lam(s) + \ell_\m(s) +1} 
+ \sum_{s \in \lam} q^{-a_\m(s) - \ell_\lam(s) -1}
&=& - \sum_{1 \leq i,j < \infty} 
\left( q^{\lam_i - \m_j + j -i} - q^{j-i} \right)~, \\
\sum_{s \in \m} q^{a_\m(s) + \ell_\lam(s) +1} 
+ \sum_{s \in \lam} q^{-a_\lam(s) - \ell_\m(s) -1}
&=& - \sum_{1 \leq i,j < \infty} 
\left( q^{\lam_i^\vee - \m_j^\vee + j -i} - q^{j-i} \right)~, \\
(q-1) \sum_i q^{\lam_i -i} &=& (q^{-1} -1) \sum_i q^{i-\lam_i^\vee}~.
\eeqa
The last equality can be rewritten as
\beq
- \sum_{i=1}^\infty q^{\lam_i -i + \frac{1}{2}} 
= \sum_{i=1}^\infty q^{-\lam_i^\vee + i - \frac{1}{2}}~,
\quad
- \sum_{i=1}^\infty q^{-i + \frac{1}{2}} 
= \sum_{i=1}^\infty q^{ i - \frac{1}{2}}~.
\eeq
This formula may be regarded as our rule of analytic continuation.
%in deriving the two lemmas.

%%%%%%% NY and NO %%%%%%%%%%%%%%%%%%%%%%%%
By using these propositions 1 and 2,
one can derive the following relations, respectively
\be
\prod_{(i,j)\in\mu } 
{
1 \over 
1 - Q\, q^{\la_i-j} t^{\mu^\vee_j-i+1}
}
\cdot
\prod_{(i,j)\in\la } 
{
1  \over 
1 - Q\, q^{-\mu_i+j-1} t^{-\la^\vee_j+i  }
}
= 
{\Pi\left(Q\, t^\rho,t^{-\rho}\right) \over 
\Pi\left(Q\, q^\la t^\rho,q^{-\mu}t^{-\rho}\right)},
\qquad Q\in{\mathbb C}
\ee
\ba
\Pi\left(Q\, q^{\la}t^{\rho},\  q^{-\mu}t^{-\rho}\right)
&=&
\left\{
\begin{array}{l}%l}
\displaystyle{
\Pi_0\left(
-Q\left({t\over q}\right)^{\ha} t^{-\la^\vee}q^{-\rho},\  q^{-\mu}t^{-\rho}
\right)
},\qquad %&|q|,|t^{-1}| < 1,
\cr
\displaystyle{
\Pi_0\left(
-Q\left({t\over q}\right)^{\ha} q^{\la}t^{\rho},\  t^{\mu^\vee}q^{\rho}
\right)
},\qquad %&|q^{-1}|,|t| < 1,
\end{array}
\right.
\label{eq:PiPitilde}
\ea
Where $\Pi(x,y)$ and $\Pi_0(x,y)$ are
the Cauchy kernel and its conjugate
in (\ref{eq:Cauchy}) and (\ref{eq:conjugateCauchy}), respectively.
These represent the equivalence 
between several expressions of the Nekrasov formula in \cite{NO} and \cite{NY1}.

%%%%%%%%%%%%%%%%%%%%%%%%%%%%%%%%%%%%%%%%%%%%%%%%%%%%%%%%%%%%

\section*{Appendix B : Formula for the Macdonald Symmetric Function}
\renewcommand{\theequation}{B.\arabic{equation}}\setcounter{equation}{0}
\renewcommand{\thesubsection}{B.\arabic{subsection}}\setcounter{subsection}{0}

%%%%%%%%%%%%%%%%%%%%%%%%%%%%%%%%%%%%%%%%%%%%%%%%%%%%%%%%%%%%

In this appendix we recapitulate basic properties of 
the Macdonald symmetric function \cite{Mac}.

%%%%%%%%%%%%%%%%%%%%%%%%%%%%%%%%%%%%%%%%%%%%%%%%%%%%%%%%%%%%%%%%%%%%%%%%%%%%

\subsection{Definition for the Macdonald Symmetric Function}

%%%%%%%%%%%%%%%%%%%%%%%%%%%%%%%%%%%%%%%%%%%%%%%%%%%%%%%%%%%%%%%%%%%%%%%%%%%%

There are various bases of the ring of symmetric functions 
in infinite number of variables $x=(x_1,x_2,\cdots)$,
for example, 
the monomial symmetric function,
the power-sum symmetric function and so on.
They are indexed by the Young diagram,
{\it i.e.}, the partition
$\la=(\la_1,\la_2,\cdots)$,
which is a sequence of non-negative integers such that
$\la_{i} \geq \la_{i+1}$ and 
$|\la | = \sum_i \la_i < \infty$.
%
%%%%%%%%%%%%%%% m %%%%%%%%%%%%%%%%
%
The monomial symmetric function $m_{\lambda}(x)$ is defined by
\be
m_{\lambda}(x)=\sum_{\sigma}
x_1^{\lambda_{\sigma(1)}}
x_2^{\lambda_{\sigma(2)}}
\cdots ,
\ee
where the summation is over all distinct permutations of 
$(\lambda_1,\lambda_2,\cdots )$.

%%%%%%%%%%%%%%% p %%%%%%%%%%%%%%%%

The power-sum symmetric function $p_{\lambda}(x)$ is defined by
\be
p_{\lambda}(x)=
p_{\lambda_1}(x)
p_{\lambda_2}(x)\cdots ,
\qquad 
p_n(x)=\sum_{i=1}^{\infty}x_i^n.
\ee
%
%%%%%%%%%% inner-product %%%%%%%%%%%%%%%%%
%
We introduce an inner-product on the ring of symmetric functions
in the following manner;
for any symmetric functions $f$ and $g$, %in power-sums $p_\la$'s,
\be
\langle f(p),\, g(p)\rangle_{q,t} 
:= f(p^*)\, g(p)\,\vert_{{\rm constant\, part}},\qquad
p_n^* := n {1-q^n \over 1-t^n} {\partial \over \partial p_n},
\ee
or equivalently
\begin{equation}
\langle p_{\lambda},p_{\mu}\rangle_{q,t} =
 \delta_{\lambda,\mu}
\prod_{r\geq1} r^{m_r} m_r ! \cdot
  \prod_{i=1}^{\ell(\lambda)}
\frac{1-q^{\lambda_i}}{1-t^{\lambda_i}},
\qquad \lambda=(1^{m_1}2^{m_2}\cdots),
\end{equation}
with $m_r\equiv \#\{i\,|\,\lambda_i=r\}$.

%%%%%%%%%% Mac %%%%%%%%%%%%%

The Macdonald symmetric function
$P_{\lambda}=P_{\lambda}(x;q,t)$
is uniquely specified by the following orthogonality and normalization,
\ba
  &&
  \langle P_{\lambda},P_{\mu}\rangle_{q,t} =0\qquad { \rm if } \;
  \lambda\neq \mu,\\
  &&
  P_{\lambda}=m_{\lambda} + \sum_{\mu<\lambda} u_{\lambda\mu}m_{\mu},
  \quad
  u_{\lambda\mu}\in {\bf Q}(q,t).
\ea
Here we used the dominance partial ordering on the Young diagrams defined as
$\lambda\geq\mu \Leftrightarrow |\lambda|=|\mu|$ and
$\lambda_1+\cdots+\lambda_i\geq\mu_1+\cdots+\mu_i$ for all $i$.

%%%%%%%%% example %%%%%%%%%%%%
The first few examples are
\ba
P_{1} &=& m_{1},\cr
  \Smat{P_{2}  \cr
        P_{1,1}}
&=&\Smat{1 & {(1+q)(1-t)\over 1-qt}  \cr
        0 & 1                   }
  \Smat{m_{2}   \cr
        m_{1,1} },\cr
  \Smat{P_{3}    \cr
        P_{2,1}  \cr
        P_{1,1,1}}
&=&\Smat{
1 
& {(1+q+q^2)(1-t)\over 1-q^2 t} 
& {(1+q)(1+q+q^2)(1-t)^2\over (1-qt)(1-q^2 t)}\cr
0 
& 1                         
& {(2+q+t+2qt)(1-t)\over 1-qt^2}            \cr
0 
& 0                         
& 1                                       }
  \Smat{m_{3}    \cr
        m_{2,1}  \cr
        m_{1,1,1}}.
\ea

%%%%%%%%%%%% limit %%%%%%%%%%%

The Macdonald symmetric function %$P_{\lambda}(x;q,t)$
contains several important symmetric functions 
as a case of special values of $q$ and $t$.
For example, 
\\
(\romannumeral1)
% When $t=q$, 
$P_{\lambda}(x;q,q) = s_{\lambda}(x)$ ; 
the Schur function.
\\
(\romannumeral2)
% When $q=0$, 
$P_{\lambda}(x;0,t) = P_{\lambda}(x;t)$ ; 
the Hall-Littlewood function.
\\
(\romannumeral3)
$\displaystyle
 \lim_{q \rightarrow 1} 
P_{\lambda}(x;q,q^{\beta}) = P^{(1/\beta)}_{\lambda}(x)$ ;
the Jack symmetric function.
\\
(\romannumeral4)
% When $t=1$, 
$P_{\lambda}(x;q,1) = m_{\lambda}(x)$ ;
the monomial symmetric function.
\\
(\romannumeral5)
% When $q=1$, 
$P_{\lambda}(x;1,t) = e_{\lambda^\vee}(x)$ ;
the elementary symmetric function.

%%%%%%%%%% scalar product %%%%%%%%%%%%%%%%%%%%%%%%%%%

The scalar-product is given by 
\be
\langle P_\la |P_\la \rangle_{q,t}
=
\prod_{s\in\la }
{
1-q^{a (s)+1} t^{  \ell(s)   }
\over 
1-q^{a (s)  } t^{  \ell (s)+1}
},
\ee
which satisfies 
\ba
\langle P_\la |P_\la \rangle_{q,t}
\langle P_{\la^\vee} |P_{\la^\vee} \rangle_{t,q} = 1,
\\
{
\left({q/ t}\right)^{|\la |\over 2}
\over 
\langle P_\la |P_\la \rangle_{q,t}
}
=
{
\left({t/ q}\right)^{|\la |\over 2}
\over 
\langle P_\la |P_\la \rangle_{q^{-1},t^{-1}}
}.
\ea
%
%%%%%%%%%%%%%% tilde P  %%%%%%%%%%%%%%%%%%
Let $\tP_{\la} (x;q,t)$ be the normalized Macdonald function
\be
\tP_{\la} (x;q,t) := 
{1 \over \langle P_\la |P_\la \rangle_{q,t}^\ha}
P_{\la} (x;q,t),
\ee
%%%%%%%%%% scalar product %%%%%%%%%%%%%%%%%%%%%%%%%%%
so that the scalar product is normalized as
$\langle \tP_\la |\tP_\la \rangle_{q,t} = 1$.

%%%%%%%%%%%%%%%%%%%%% skew Mac %%%%%%%%%%%%%%%%%%%

The normalized skew-Macdonald symmetric function 
$\tP_{\la/\mu}(x;q,t)$ is defined by
\be
\tP_{\la/\mu}(x;q,t) 
:=
\tP_\mu^*\left(x;q,t\right) \, \tP_\la(x;q,t), 
\ee
where $*$ is acted on the power-sum as
$p_n^* := n {1-q^n \over 1-t^n} {\partial \over \partial p_n}$.
The relation with the usual skew-Macdonald function $P_{\la/\mu} (x;q,t)$ is
\be
\tP_{\la/\mu} (x;q,t) := 
{\langle P_\mu |P_\mu \rangle_{q,t}^\ha \over 
\langle P_\la |P_\la \rangle_{q,t}^\ha}
P_{\la/\mu} (x;q,t).
\ee
Finally let $\iota\tP_{\la/\mu}(x;q,t)$ be 
the skew-Macdonald function with the involution $\iota$ 
acting on the power-sum $p_n$ as $\iota(p_n) = -p_n$.

%%%%%%%%%%%%%%%%%%%%%%%%%%%%%%%%%%%%%%%%%%%%%%%%%%%%%%%%%%%%%%%%%%%%%%%%%%%%

\subsection{%Basic Properties : 
Symmetries and Cauchy Formulas}

%%%%%%%%%%%%%%%%%%%%%%%%%%%%%%%%%%%%%%%%%%%%%%%%%%%%%%%%%%%%%%%%%%%%%%%%%%%%

%Next we turn to show the basic properties of the (skew) Macdonald symmetric function.
%
%%%%%%%%%% symmetry %%%%%%%%%%%%%%%%%%%%%
The skew-Macdonald symmetric function enjoys the following symmetries
\ba
\tP_{\la/\mu} (Qx; q,t) 
&=& 
Q^{|\la |-|\mu|}
\tP_{\la/\mu} (x; q,t),
\\
\tP_{\la/\mu} \left(x; q^{-1},t^{-1}\right) 
&=&
\left({q\over t}\right)^{ {|\la |-|\mu| \over 2} }
\tP_{\la/\mu} (x; q,t),
\\
\tP_{\la^\vee/\mu^\vee} (x; t,q) 
&=& 
\omega_{q,t} \tP_{\la/\mu} (x; q,t), 
\label{eq:skewConjugate}
\ea
with the endmorphism $\omega_{q,t}$ such that
\be
\omega_{q,t} (p_n) 
=
(-1)^{n-1}{1-q^n \over 1-t^n} p_n.
\ee
Note that 
\be
\tP_{\la/\mu} (x; q,t) 
\tP_{\la^\vee/\mu^\vee} (y; t,q) 
=
P_{\la/\mu} (x; q,t) 
P_{\la^\vee/\mu^\vee} (y; t,q). 
\ee
%
%%%%%%%%%%%%%%%%%%%%%%%%%%%%%%%%%%%%%%%%%%%%%%%
%
In the $t=q$ case, the Schur function satisfies also
\be
s_{\la^\vee}(x) = \iota s_\la(-x) = (-1)^{|\la |}\iota s_\la(x).
\ee
%however, the Macdonald function does not have.

%%%%%%%%%% Cauchy %%%%%%%%%%%%%%%%%%%%%%%%%%%
The following Cauchy formula is especially important;
\ba
\sum_\la 
\tP_\la (x;q,t) \tP_\la (y;q,t)
&=&
\Pi(x,y),
\cr
&\!\!\! := \!\!\!&
\prod_{k\geq 0}
\prod_{i,j}
{1-t x_i y_j q^k 
\over 
 1-  x_i y_j q^k },
\qquad |q|<1,
\cr
&=&
\exp\left\{
\sum_{n>0}{1\over n}{1-t^n \over 1-q^n} p_n(x) p_n(y)
\right\}.
\label{eq:Cauchy}
\ea
By acting %If we act 
on the variables $y$ with the endmorphism $\omega_{q,t}$ %, then 
we get %the following conjugate formula
\ba
\sum_\la 
\tP_\la (x;q,t) \tP_{\la^\vee} (y;t,q)
&=&
\Pi_0(x,y),
\cr
&\!\!\! := \!\!\!&
\prod_{i,j}(1+ x_i y_j),
\cr
&=&
\exp\left\{
\sum_{n>0}{(-1)^{n-1}\over n} p_n(x) p_n(y)
\right\}.
\label{eq:conjugateCauchy}
\ea
%
%%%%%%%%%% involution %%%%%%%%%%%
%
If we act  with the involution $\iota$,
then $\Pi(x,y)$ and $\Pi_0(x,y)$ are mapped to their inverse, 
\ba
\sum_\la 
\tP_\la (x;q,t)\ \iota \tP_\la (y;q,t)
&=&
\Pi(x,y)^{-1},
\\
\sum_\la 
\tP_\la (x;q,t)\ \iota \tP_{\la^\vee} (y;t,q)
&=&
\Pi_0(x,y)^{-1}.
\ea

%%%%%%%%%%%%%% Cauchy %%%%%%%%%%%%%%

The Cauchy formulas for the skew-Macdonald function are
 \ba
\sum_\la 
 \tP_{\la/\mu} (x;q,t) \tP_{\la/\nu} (y;q,t)
&=&
\Pi(x,y)
\sum_\la 
\tP_{\mu/\la} (y;q,t) \tP_{\nu/\la} (x;q,t),
\\
\sum_\la 
\tP_{\la/\mu} (x;q,t) \tP_{\la^\vee/\nu^\vee} (y;t,q)
&=&
\Pi_0(x,y)
\sum_\la 
\tP_{\mu^\vee/\la^\vee} (y;t,q) \tP_{\nu/\la} (x;q,t).
\ea
Using this formula successively, we obtain the following trace formula
\\
{\bf Proposition.}
{\it For} $|a|,|b|<1$,
\ba
&\!\!\! \displaystyle{\sum_{\la,\mu,\nu,\rho} } \!\!\!&
a^{|\la |} b^{|\nu |} 
\tP_{\la/\mu} (x;q,t) \tP_{\nu/\mu} (y;q,t)
\tP_{\nu/\rho} (z;q,t) \tP_{\la/\rho} (w;q,t)
\cr
&=&
\prod_{k> 0} 
\Pi(a^{k} b^{k} x,y) \,
\Pi(a^{k} b^{k} z,w) \,
\Pi(a^{k} b^{k-1} x,w) \, 
\Pi(a^{k-1} b^{k} z,y) \,
(1-a^k b^k)^{-1} .
\ea

%%%%%%%%%%%%%%%%%%%%%%%%%%%%%%%%%%%%%%%%%%%%%%%%%%%%%%%%%%%%%%%%%%%%%%%%%%%%

\subsection{Specialization Formulas} %Special variable case}

%%%%%%%%%%%%%%%%%%%%%%%%%%%%%%%%%%%%%%%%%%%%%%%%%%%%%%%%%%%%%%%%%%%%%%%%%%%%

Here we give the specialization formulas for the special variables.
In the case of $|t|>1$,
one can set the variables $x$ to the principal specialization
$x = t^\rho$ that stands for $x_i = t^{\ha -i}$.
%
%%%%%%%%%%%%%%%% list %%%%%%%%%%%%%%%%%%%%%%%%%%%
The following are the formulas for the principal specialization  \cite{Mac}
\ba
P_\la(t^{\rho};q,t)
&=&
\prod_{s\in\la} 
{-t^{\ha} q^{a(s)}
\over 
1-q^{a(s)} t^{\ell(s)+1} 
},
\qquad
\iota P_\la(t^{\rho};q,t)
=
\prod_{s\in\la} 
{t^{\ha} t^{\ell(s)}
\over 
1-q^{a(s)} t^{\ell(s)+1} 
},
\quad |t^{-1}|<1,
\cr
P_\la(t^{-\rho};q,t)
&=&
\prod_{s\in\la} 
{t^{\ha} t^{\ell(s)}
\over 
1-q^{a(s)} t^{\ell(s)+1} 
},
\qquad
\iota P_\la(t^{-\rho};q,t)
=
\prod_{s\in\la} 
{-t^{\ha} q^{a(s)}
\over 
1-q^{a(s)} t^{\ell(s)+1} 
},
\quad |t|<1.
\cr
&\!\!\!\!\!\!&
\ea
%%%%%%%%%% t.q %%%%%%%%%%
\ba
P_{\la^\vee}(q^{\rho};t,q)
&=&
\prod_{s\in\la} 
{q^{-\ha} q^{-a(s)}
\over 
1-q^{-a(s)-1} t^{-\ell(s)} 
},
\quad
\iota P_{\la^\vee}(q^{\rho};t,q)
=
\prod_{s\in\la} 
{-q^{-\ha} t^{-\ell(s)}
\over 
1-q^{-a(s)-1} t^{-\ell(s)} 
},
\quad |q^{-1}|<1,
\cr
P_{\la^\vee}(q^{-\rho};t,q)
&=&
\prod_{s\in\la} 
{-q^{-\ha} t^{-\ell(s)}
\over 
1-q^{-a(s)-1} t^{-\ell(s)} 
},
\quad
\iota P_{\la^\vee}(q^{-\rho};t,q)
=
\prod_{s\in\la} 
{q^{-\ha} q^{-a(s)}
\over 
1-q^{-a(s)-1} t^{-\ell(s)} 
},
\quad |q|<1.
\cr
&\!\!\!\!\!\!&
\ea

%%%%%%%%%%%%%% proposition %%%%%%%%%%%%%%

By using the formula of the analytic continuation
(\ref{eq:AC}), we have
\\
{\bf Proposition.}
\be
\tP_{\la^\vee/\mu^\vee}\left(t^{\pm\eta^\vee}q^{\pm\rho}; t,q\right)
=
(-1)^{|\la |-|\mu|}
\left({q\over t}\right)^{|\la |-|\mu| \over 2}
\tP_{\la /\mu } \left(q^{\mp\eta}t^{\mp\rho}; q,t\right),
\qquad
|q^{\mp 1}|, |t^{\pm 1}| < 1.
\label{eq:ACMac}
\ee
Here $q^\eta t^\rho$ stands for 
$x_i = q^{\eta_i} t^{\ha -i}$.
%
%%%%%%%%%%% principal case %%%%%%%%%%%%%%%%%%%%%
In the principal case
\ba
\tP_{\la^\vee}\left(q^{\rho}; t,q\right)
&=&
\tP_\la \left(t^{-\rho}; q,t\right)
(-1)^{|\la |} \left({q\over t}\right)^{|\la | \over 2},
\\
&=&
\iota \tP_\la \left(t^{\rho}; q,t\right)
(-1)^{|\la |} \left({q\over t}\right)^{|\la | \over 2},
\\
&=&
\tP_\la \left(t^{\rho}; q,t\right)
 \left({q\over t}\right)^{|\la | \over 2}
 \prod_{s\in\la} q^{-a(s)} t^{\ell(s)}.
\label{eq:ACprincipalMac}
\ea

%%%%%%%%%%%%%%%%%%%%%%%%%%%%%%%%%%%%%%%%%%%%%%%%%%%%%%%%%%%%

\end{document}